# Lasing of Quantum-Dot Micropillar Lasers under Elevated Temperatures

Andrey Babichev, Ivan Makhov, Natalia Kryzhanovskaya, Alexey Blokhin, Yuriy Zadiranov, Yulia Salii, Marina Kulagina, Mikhail Bobrov, Alexey Vasil'ev, Sergey Blokhin, Nikolay Maleev, Maria Tchernycheva, Leonid Karachinsky, Innokenty Novikov, and Anton Egorov

*Abstract—* A comprehensive numerical modelling of microcavity parameters for micropillar lasers with optical pumping was presented. The structure with a hybrid dielectric-semiconductor top mirror has a significantly higher calculated quality-factor (~65000 for 5 µm pillar) due to better vertical mode confinement. The minimum laser threshold (~370 µW for 5 µm pillar) coincided with a temperature of 130 K, which is close to zero gain to cavity detuning. Lasing up to 220 K was demonstrated with a laser threshold of about 2.2 mW.

## I. INTRODUCTION

Optical pumping of quantum dots (QDs) located inside the micropillar cavity is of interest for various applications (i.e. photonic quantum information technologies, free-space communication, integrated photonics, quantum nanophotonics and neuromorphic computing [1–4]). As a result, high-performance quantum single-photon light sources [5], two-level emitters [6], thresholdless lasers [1, 7, 8], topological lasers [9], bimodal lasers [10] as well as highly homogeneous arrays of vertically emitting microlasers [4] were made of a QD micropillar cavity.

Recently, photonic reservoir computing (RC) consisted of vertical-cavity surface-emitting lasers (VCSELs) nodes was demonstrated [11]. The coupling scheme could be expanded to many more nodes, if QD micropillar lasers replace the VCSELs array [11]. At the same time, new challenge arises such as operation at cryogenic temperatures of micropillar lasers [2, 4, 12]. Indeed, the maximal lasing temperature of a QD micropillar lasers was about 130 K [2].

Herein, the results of modelling and experimental study of a QD micropillar cavity lasers with non-absorbing hybrid semiconductor-dielectric top mirror was presented, demonstrating lasing under elevated temperatures of up to 220 K.

## II. MICROPILLAR STRUCTURE: MODELLING AND FABRICATION

Firstly, we discussed the design of a micropillar cavity with semiconductor mirrors and then with a dielectric-semiconductor hybrid top mirror. Numerical modeling was performed within the time domain finite difference method (FDTD) [13]. To simplify the calculations, absorption in the gain region, as well as absorption on free charge carriers, were not considered. A probe box was formed around the microcavity, along the edges of which the proportions of optical radiation emitting from the microcavity into the upper and lower hemispheres, as well as through the surface of the microcavity sidewalls, were determined. The parameters of the computational grid were chosen in a such a way as to minimize the change in the calculated characteristics of the microcavities with a further decrease in the grid step.

The modal composition of the microcavity emission was determined using a discrete Fourier transform of the microcavity response to the emission of a microcavity-centered dipole with a wide emission spectrum. The X-Y distribution of the electromagnetic field inside the microcavity was analyzed at the corresponding resonant frequencies to determine the type of optical mode. At the resonant frequency of the fundamental mode, the time domain response to a short dipole excitation pulse was calculated. The far field patterns were modelled based on the near field patterns.

Section II,*A* presents 1D numerical simulation of the planar GaAs 1$\lambda$-cavity with different number of pairs of the distributed Bragg reflector (DBR) and their composition. The case of a cold cavity was considered (without considering absorption in the gain region). Sections II,*B*–*D* present the results of three–dimensional (3D) simulations of a cylindrical 1$\lambda$-cavity designed to minimize mode volume, based on 35 and 25 pairs of $Al_{0.2}Ga_{0.8}As/Al_{0.9}Ga_{0.1}As$ layers in the bottom and top DBRs. The Q-factor, mode volume, wavelength, and photon extraction efficiency for the different numerical apertures were estimated for the case of the fundamental vertical mode.

Manuscript received 19 July 2024; revised XX XXXX 2024; accepted XX XXXX 2024. Date of publication XX XXXX 2024; date of current version XX XXXX 2024. The authors from Ioffe Institute acknowledge support in part by the grant of the Russian Science Foundation No. 22-19-00221, https://rscf.ru/project/22-19-00221/ for the structure design, MBE epitaxy, and the study of photoluminescence and lasing spectra. N. Kryzhanovskaya thanks the Basic Research Program at the HSE University for support of the study of photoluminescence spectra measured at 90° tilted sample of microcavity structure. *(Corresponding authors: Andrey Babichev.)*

Andrey Babichev, Alexey Blokhin, Yuriy Zadiranov, Yulia Salii, Marina Kulagina, Mikhail Bobrov, Alexey Vasil'ev, Sergey Blokhin, Nikolay Maleev and Anton Egorov are with Ioffe Institute, 194021 Saint Petersburg, Russia (e-mail: a.babichev@mail.ioffe.ru);

Ivan Makhov, and Natalia Kryzhanovskaya are with HSE University, 190008 Saint Petersburg, Russia;

Maria Tchernycheva is with Centre de Nanosciences et Nanotechnologies, Université Paris-Saclay, CNRS, UMR9001, 91120 Palaiseau, France;

Leonid Karachinsky, Innokenty Novikov, and Anton Egorov are with ITMO University, 197101 Saint Petersburg, Russia.

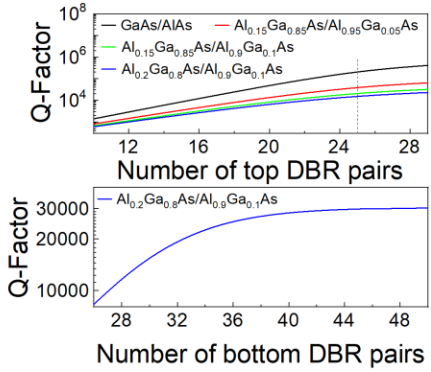

**Fig. 1.** (a) Dependence of the Q-factor of a planar 1$\lambda$-microcavity on the number of pairs in the top DBR with a change in the layers composition. (b) Dependence of the Q-factor of a planar 1$\lambda$-microcavity on the number of pairs in the bottom DBR.

*A. 1D numerical simulation of the planar microcavity Q-factor depending on the number of pairs of the top DBR*

Figure 1,*a* shows the results of modeling the Q-factor of 1$\lambda$-cavity with different aluminum content in the top DBR layers. The pairs number in the bottom DBR (of the same composition) was fixed on 30. The highest Q-factor was provided by the combination of AlAs/GaAs layers due to the maximum refractive index contrast. At the same time, the use of AlAs or $Al_{0.95}Ga_{0.05}As$ in the DBR is limited due to the oxidation of these layers in the ambient atmosphere, which leads not only to low mechanical reliability of the cavity, but also to uncontrolled fluctuations in cavity parameters. In addition, the use of GaAs in the top DBR leads to absorption losses in the top mirror during non-resonant pumping into the GaAs matrix [2].

As the aluminium (Al) content in the high refractive index layer increases and the Al content in the low refractive index layer decreases, a drop in the Q-factor was observed at the fixed pairs in the top DBR. An increase in the number of pairs of top DBR leads to the enlarge in Q-factor with a tendency to saturate the behavior when the pairs number of top DBRs is equal to the pairs number of bottom DBR.

The use of $Al_{0.2}Ga_{0.8}As$ layers in the top DBR allows minimizing light absorption and increasing pumping efficiency when using optical pumping at a wavelength in the range of 700–820 nm [2]. Figure 1,*b* shows the results of the Q-factor modelling for a 1$\lambda$-cavity with a change in the number of pairs in the bottom $Al_{0.2}Ga_{0.8}As/Al_{0.9}Ga_{0.1}As$ DBR. The number of pairs in the top DBR of the same composition was fixed at 25, which is due to the need to ensure efficient radiation extraction through the top mirror.

An increase in the pairs number in the bottom DBR leads to an increase in the Q-factor to $30\cdot10^3$ with an abrupt saturation of the Q-factor compared to the number of pair in the bottom DBR above 35. In fact, a simultaneous increase in the number of pairs in the bottom and top DBRs yields to a further increase in the Q-factor. However, the use of many pairs in the bottom and top mirrors was associated with the need to increase the pillar etching depth, which could negatively affect both the stability of the etching hardmask and the verticality of the pillar sidewalls. Moreover, considering the use of $Al_{0.2}Ga_{0.8}As$, an increase in the roughness of the layers is possible.

*B. 3D numerical simulation of the microcavity parameters depending on the pillar diameter*

The modelled pillar has a maximum etching depth (i.e., to the substrate) and no tilt when etching the sidewalls. For a pillar diameter greater than 10 μm, the wavelength of vertical mode corresponds to the resonance wavelength of the planar cavity structure (cf. Figure 2,*a*). As the pillar diameter decreased, a blueshift in the mode wavelength was observed, which was clearly observed when the pillar diameter was less than 2 μm.

A decrease in the fundamental vertical mode volume ($V_m$) was observed with reduction pillar diameter, and the Q-factor remains virtually unchanged down to 2 μm (Figure 2,*a*). A further decrease in the pillar diameter leads to an abrupt drop in the Q-factor (from $2.5\cdot10^4$ to $1.3\cdot10^4$).

Figure 2,*b* displays the effect of the pillar diameter on the photon–extraction efficiency, PEE (in the upper hemisphere), the sidewalls leakage efficiency, SLE, and the bottom leakage efficiency, BLE (into the substrate, in the lower hemisphere) for the fundamental vertical mode. As the lateral size of the microcavity decreased, the overall (i.e., numerical aperture, NA = 1) photon–extraction efficiency increased from ~67% to ~75%, mainly due to a decrease in the BLE value (from ~30% to ~21 %), probably due to improved lateral optical confinement.

For the pillar diameters less than 2 μm, a drop in the overall efficiency was observed. In this case, an abrupt increase in radiation leakage into the substrate has been demonstrated (enlarge in BLE value from ~21% to ~48%). In addition, radiation leakage through the surface sidewalls is shown (an increase in SLE value from ~(1–2)% to ~7%), which is associated with light scattering at the air-semiconductor interface (cf. inserts to Figure 2,*b*).

The dependences of the PEE on the pillar diameter for various numerical apertures are presented in Figure 2,*c*. At small NA values, an almost uniform decrease in the overall efficiency was observed as the lateral size of the microcavity decreased. At the same time, with an increase in the radiation collection area (i.e., an increase in the NA value), an abrupt rise in the photon–extraction efficiency occurs with a pronounced maximum for a certain range of pillar diameters. This behavior is due to a nonlinear change in the divergence diagram of the output radiation of a cylindrical microcavity with a decrease in its diameter (cf. insets to Figure 2,*c*). In addition, an increase in leakage into the substrate (BLE value) and sidewalls leakage (SLE value) at small sizes has also been demonstrated.

To summarize, the pillar diameter of 3–5 μm is of most interest since optics with a numerical aperture of about 0.4 are predominantly used.

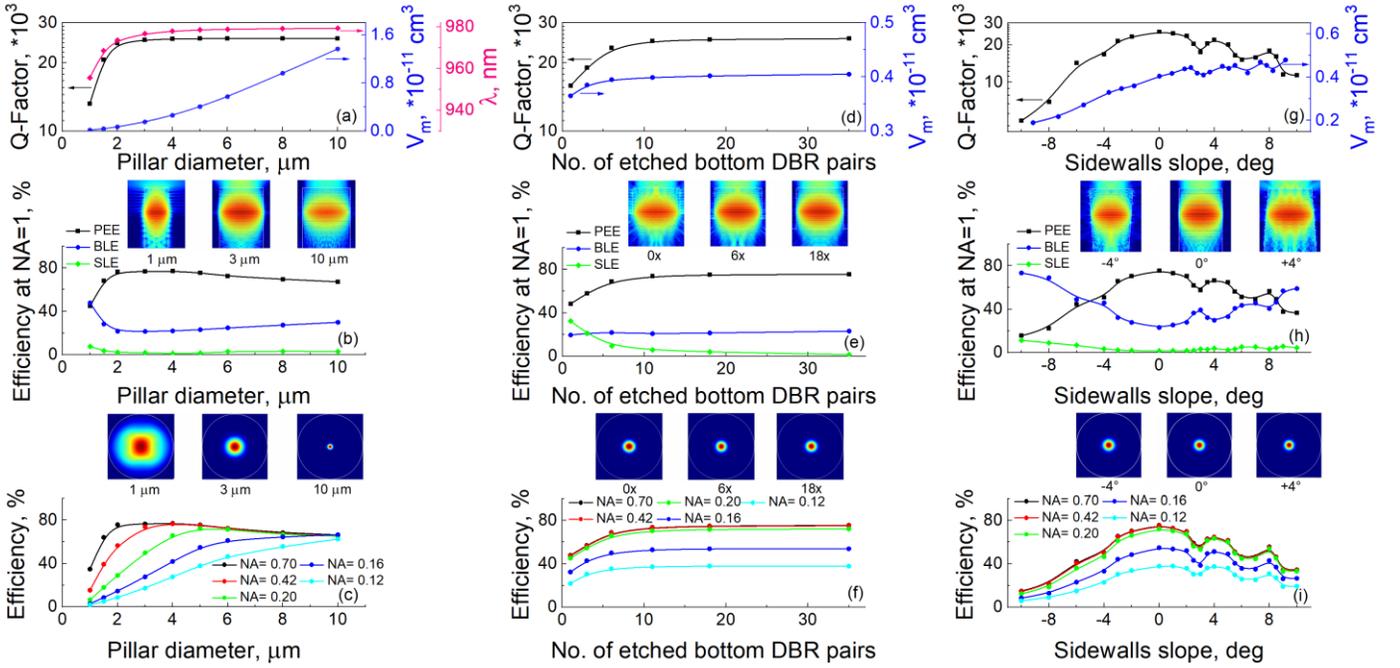

**Fig. 2.** The influence of the pillar diameter on: (a) the wavelength of the fundamental vertical mode (λ value), Q-factor and mode volume; (b) PPE, SLE, and BLE values at NA=1; (c) PPE values at different NA values. Insets (*b* panel): the distributions of the electromagnetic field (*E*-field) of the fundamental mode in the X-Z plane for microcavities with diameters of 1, 3 and 10 μm. Insets (*c* panel): the angular distributions of radiation intensity in the far field for microcavities with diameters of 1, 3, and 10 μm; The influence of the number (No.) of etched bottom DBR pairs on: (d) Q-factor and mode volume; (e) PPE, SLE, and VLE values at NA=1; (f) PPE values at different NA values. Insets (*e* panel): the distributions of the E-field of the fundamental mode in the X-Z plane for microcavities with the bottom DBR etching depth of 0, 6 and 18 pairs. Insets (*f* panel): the angular distributions of radiation intensity in the far field for microcavities with the bottom DBR etching depth of 0, 6 and 18 pairs; The influence of the micropillar sidewalls tilt on: (g) Q-factor and mode volume; (h) PPE, SLE, and VLE values at NA=1; (i) PPE values at different NA values. Insets (*h* panel): the distributions of the E-field of the fundamental mode in the X-Z plane for micropillars with sidewalls tilt –4, 0 and +4 degrees. Insets (*i* panel): the angular distributions of radiation intensity in the far field for micropillars with sidewalls tilt –4, 0 and +4 degrees.

### C. 3D numerical simulation of the microcavity parameters depending on the etching depth of the bottom DBR

For the 5 μm pillar, an increase in the Q-factor was observed from $1.6·10^4$ to $2.5·10^4$ with an increase in the etching depth of the bottom DBR (cf. Figure 2,*d*), accompanied by a slight increase in the mode volume. In addition, a significant decrease in radiation leakage through the side surface was demonstrated (cf. Figure 3,*e*). In fact, the SLE value drops from 30% to ~2% due to the weakening of the light scattering effect at the air-semiconductor interface (cf. insets to Figure 3,*e*), which leads to an increase in the overall photon extraction efficiency from 48% to 75%.

For the case of different numerical apertures, an increase in the etching depth leads to an abrupt rise in the photon extraction efficiency and saturation of the dependence at an etching depth of more than 10 pairs (cf. Figure 3,*f*). The photon extraction efficiency of more than 70% was realized at a NA value of more than 0.2 due to the narrow radiation divergence diagram (cf. insets to Figure 3,*f*).

To summarize, the etching depth of the bottom DBR does not affect on the microcavity parameters when the etching depth is more than 20 pairs, obviously exceeding the penetration depth of the *E*-field of the fundamental vertical mode into the bottom DBR, which reduces the requirements for the etching depth of the DBR.

### D. 3D numerical simulation of the microcavity parameters depending on the micropillar sidewalls tilt

It has been shown that any deviation of the sidewalls tilt from the normal yields to a drop in the Q-factor of the 5 μm pillar (cf. Figure 2,*g*), accompanied by fluctuations due to interference. With a negative sidewalls tilt, a decrease in the mode volume was observed due to an effective reduction in the microcavity size, and with positive sidewalls tilt, an increase in the mode volume due to an effective enlarge in the microcavity size.

With a positive sidewalls tilt, a decrease in the overall photon–extraction efficiency was observed (cf. Figure 4,*h*) due to an increase in radiation leakage into the lower hemisphere (an increase in the BLE value). As the negative sidewalls tilt increased, a rapid enlarge in radiation leakage into the substrate (increase in the BLE value) was observed, accompanied by an increase in the sidewalls leakage (increase in the SLE value). It should be noted that as the sidewalls tilt rised, hybridization of the fundamental mode with higher order modes was observed, presumably due to light scattering at the inclined boundary and interference.

The behavior of the photon extraction efficiency depending on the sidewalls tilt for the case of different numerical apertures (cf. Figure 4,*i*) is like the behavior of the overall photon extraction efficiency discussed above. With a positive sidewalls tilt, a slight drop in the radiation divergence diagram was observed (cf. insets to Figure 4,*i*), which is apparently due to an effective increase of the micropillar size. A negative sidewalls tilt leads to an increase in the radiation divergence diagram, which is apparently associated with an effective reduction of the micropillar size.

To summarize, the microcavity shape acts as an antenna, redirecting the radiation. However, when the sidewalls were tilted no more than ±2 degree, a relatively weak change in the microcavity characteristics was observed.

*E. 3D numerical simulation of the Q-factor depending on the top DBR pairs*

First, the simulation was performed on a 5 μm pillar for three different structure designs of 1λ-cavity. The first cavity structure had 33/29 pairs of $Al_{0.2}Ga_{0.8}As/Al_{0.9}Ga_{0.1}As$ forming DBRs [12]. The second one was based on 35/27 pairs of $Al_{0.2}Ga_{0.8}As/Al_{0.9}Ga_{0.1}As$ forming DBRs. The third hybrid structure had 35/27 pairs of $Al_{0.2}Ga_{0.8}As/Al_{0.9}Ga_{0.1}As$ forming DBRs with additional two λ/4*n*-thick dielectric pairs ($SiO_2/Ta_2O_5$ pairs) located on the top of semiconductor mirror.

At room temperature (300 K), the calculated value of the Q-factor for the first design (~48300) exceeds the similar value for the second micropillar cavity (~35500). At the same time, the third hybrid structure demonstrates a significantly higher Q-factor (~76800) due to the high refractive index contrast of dielectric pairs (~0.61 [14]).

Reducing the temperature yields the Q-factor for the third design around 65000 and 58600 at 130 K and 10 K, respectively. The change to a semiconductor top mirror (the second design) leads to the Q-factor of about 31700 and 30600 at 130 K and 10 K, respectively. Increasing the pairs of the output mirror (the first design) results the Q-factor of about 40700 and 36600 at 130 K and 10 K, respectively.

To summarize, the use of structure design with a hybrid non-absorbing top mirror is more promising if it is planned to significantly increase the Q-factor of micropillar cavity. The latter makes it possible to reduce the laser threshold and implement lasing under elevated temperatures.

III. MICROPILLARS FABRICATION

The structure was grown by molecular-beam epitaxy (MBE). One-λ thick GaAs cavity was fabricated between the bottom and top mirrors, consisting of 35 and 27 non-absorbing $Al_{0.2}Ga_{0.8}As/Al_{0.9}Ga_{0.1}As$ layers with a thickness of λ/4*n*. Three layers of self-assembled Stranski-Krastanow QDs formed from $In_{0.5}Ga_{0.5}As$ layers (5.5 monolayer thick) were used as gain region. To mitigate the vertical stacking of QDs, GaAs barriers with a thickness of 20 nm were used. The thickness of the GaAs absorbing layer was about 210 nm.

Optimization of growth conditions made it possible to reduce the root-mean-square roughness of the microcavity structure surface to 0.3 nm. The Q-factor of the planar structure, determined at 300 K from a high-resolution reflectance spectrum is about 4100. The difference in the resonance energy ($E_0$) shift along the radius of the 3-inch wafer is only 8 meV.

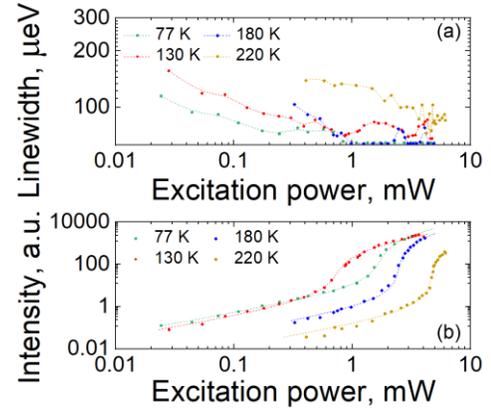

**Fig. 3.** (a) Spectral linewidth versus excitation power for 5 μm pillar determined at different temperatures (77 K, 130 K, 180 K, and 220 K). (b) Excitation power-dependent input-output characteristics measured at different temperatures.

Micropillars were fabricated using photolithography and ion-beam etching. The conical shaped mesas were fabricated by micropillar etching (i.e., positive sidewalls tilt) with an estimated tilt less than 1 degree. To fabricate the hybrid micropillar cavity structure, two additional quarter-wave $SiO_2/Ta_2O_5$ pairs were deposited on the top semiconductor DBR by magnetron sputtering process.

The samples were mounted on a copper holder in a closed-cycle optical cryostat (Montana Instruments Cryostation s50) with temperature control over the range of 5 – 300 K. To measure the emission spectra, an 808 nm semiconductor laser diode operating in continuous-wave mode was used. A 100x, NA = 0.42 microobjective (Mitutoyo M Plan Apo NIR) was used to both excite and collect photoluminescence (PL) emission. The focused spot size of laser beam on the pillar surface was about 1.0–2.0 μm. A monochromator (Andor Shamrock 500i grating) with 500 mm focal length, equipped with a thermoelectrically cooled back-illuminated silicon CCD matrix (Andor DU 401A BVF), was used to collect emission. The spectral resolution of the monochromator with a 1200 lines/mm grating was about 0.05 nm (~ 64 μeV at the 981 nm line).

IV. EXPERIMENTAL RESULTS

*A. The power-dependent input–output characteristics for 5 μm pillar*

The power-dependent input–output characteristics (double logarithmic scale), measured at various temperatures, are presented on Figure 3. Pseudo–Voigt line-shapes fitting is used to determine the integrated intensities as well as the spectral linewidth of the fundamental cavity mode [12, 15–19]. Clear S-curves as well as reduced spectral linewidths confirm about the laser transition at temperatures up to 220 K.

The coupling between the lasing mode and the QD gain inside high-quality micropillar cavity, called the *β* factor [2]. Micropillar lasers with moderate spontaneous emission factor was used to realize energy-efficient RC [4]. In fact, a high $\beta$-factor leads to the partial injection locking [20] that can be compensated by increasing the excitation power (decreasing the energy efficiency [4]).

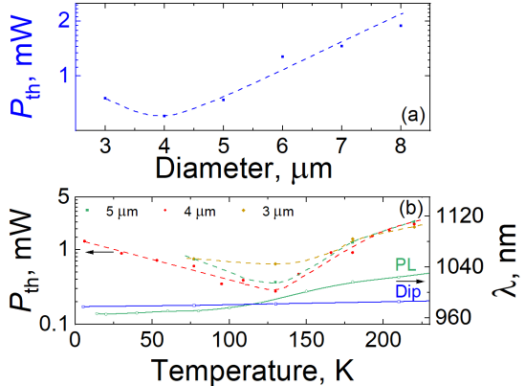

**Fig. 4.** (a) The laser thresholds for pillars of different diameters measured at 77 K. (b) Temperature dependence of the laser threshold for pillars of different diameters (left Y axis). Reflectance dip and PL positions at different temperatures (right Y axis).

To determine the $\beta$ factor, the rate-equation fitting [12] was used (cf. Figure 3,b). Absorbed power, $P_{abs}= h\nu^L\gamma[n/(1 + n)(1+\xi)(1 + \beta n) − \xi\beta n]$, where $\nu^L$ is the fundamental mode frequency, $\gamma$ is the cavity decay rate, $n$ is the intracavity photon number and $h$ is the Planck's constant, was determined from the excitation power divided by the photon effective pump absorption factor, $\kappa$.

About 0.5 % of the spontaneous emission factor is determined at 77 K. Increasing the temperature to 130 K leads to an enlarge in the $\beta$ factor to 0.7 %. A further increase in temperature to 220 K leads to a decrease in the spontaneous emission factor to about 0.05%.

The extracted laser threshold ($P_{th}$), as mentioned previously [12], was about 0.74 mW at 77 K. The Q-factor of about 19300 with the fundamental emission mode (981 nm) is derived at the $P_{th}$ value (bare Q-factor). A further increase in the excitation power (over 1 mW) gives a Q-factor of at least 19600, limited by the monochromator spectral resolution. As a result, the use of a hybrid top mirror improves the Q-factor relative to the structure with semiconductor mirrors, which coincides with the simulation results discussed above in Section II.

The minimum laser threshold (~370 µW) corresponds to a temperature of 130 K with a bare Q-factor of about 15000. This threshold power is significantly lower (~4 times) than previously reported value ($P_{th}$ ~ 1.5 mW at 130 K [2]). In fact, the determined power conversion efficiency, PCE is about 14 % (for 808 nm excitation power) versus a PCE value of ~ 3.5% for a cavity structure with 27/23 pairs GaAs/Al$_{0.9}$Ga$_{0.1}$As DBRs [2].

Increasing the temperature to 180 K gives a laser threshold of 1.2 mW with a bare Q-factor limited by the monochromator resolution. Lasing at elevated temperature (220 K) leads to an increase in the threshold excitation power to 2.3 mW while reducing the bare Q-factor by approximately 12600.

The power conversion efficiency allows us to determine the threshold absorbed power [2], which is about 96 µW and 52 µW at 77 K and 130 K for the studied 5 µm pillar lasers with an active region based on three QDs layers. Previously, a lower lasing threshold (~30 µW at 77 K) was demonstrated for a laser based on a micropillar cavity with 33/29 pairs Al$_{0.2}$Ga$_{0.8}$As/Al$_{0.9}$Ga$_{0.1}$As DBRs and single layer of QDs in active region [12].

The implementation of RC based on a micropillars involves the diffractively coupling in laser array via injection locking, but thermal or nonlinear effects influence the extrapolated dependence of the available injection locking range [4]. As a result, the excitation power of micropillar lasers with absorbing GaAs/Al$_{0.9}$Ga$_{0.1}$As mirrors in injection locking experiments is limited to about 0.5 mW [4]. In contrast, the use of non-absorbing mirrors minimizes the heating effect and enlarges the PCE value [12]. As a result, it is possible to increase the excitation power of the micropillar lasers used for injection locking, which make it possible to reduce the number of micropillars in the array to achieve the same locking range.

Previously, the temperature stability of micropillar cavity lasers with non-absorbing semiconductor mirrors was assessed at 77 K [12]. A small blue energy shift ($\Delta E$) was demonstrated for 5.4 µm pillar ($\Delta E$=60 µeV at 6×$P_{th}$) [12]). Further increases in excitation power was primarily related due to thermal effect and resulted a slight redshift ($\Delta E$ =47 µeV at 10×$P_{th}$).

Herein, using a hybrid non-absorbing top DBR, we were demonstrated a similar fine blue energy shift up to roll-over of $\Delta E(P_{exc})$ dependence ($\Delta E$=45 µeV at 6×$P_{th}$). Moreover, increase the temperature to 130 K yields of approximately 40 µeV blue energy shift up to rollover of $\Delta E(P_{exc})$ dependence (1.1 mW). A small redshift ($\Delta E$ =(-9) µeV) is observed at ten times the laser threshold (3.7 mW).

*B. Temperature study of micropillar lasers of various diameters*

For micropillars with a diameter of 3–8 µm at 77 K, clear *S*-curves of the power-dependent input–output characteristics are demonstrated, as well as reduced spectral linewidths. The extracted value of laser thresholds are presented in Figure 4,*a*. The minimum $P_{th}$ value (~0.6 mW) corresponds to 4 µm pillar. Similar laser threshold (~0.75 mW) is reported for 3 and 5 µm pillars. The maximal $\beta$ factor is about 1.8 % for 3 µm pillar laser (due to the smaller mode volume) and drops to 0.06 % as the pillar diameter increases to 8 µm.

Clear signaturing of the laser transition is demonstrated up to 220 K. The minimum laser threshold values correspond to temperature of about 130 K for 3–5 µm pillars (cf. Figure 4,*b*). The lowest threshold value is about 280 µW (4 µm pillar). Increasing the temperature to 220 K leads to an enlarge in the laser threshold to 2.2 mW. To evaluate the temperature behavior of the laser threshold, the resonance wavelength (reflectance dip) and PL peak positions of the planar cavity structure were studied (Figure 4,*b*). It is shown that the zero gain to cavity detuning value corresponds to a temperature of about 115 K, which correlates with the position of the minimum laser threshold.

## V. Conclusion

The results of 3D modeling of a micropillar cavity designed to minimize the mode volume were presented. It was shown that 3–5 μm pillar is of most interest since optics with a numerical aperture of about 0.4 are predominantly used. When the etching depth of the bottom mirror is more than 20 pairs and the side walls are tilted less than ±2 degrees, the microcavity parameters do not change.

Clear saturation of the Q-factor is demonstrated with an increase in the number of pairs in the bottom semiconductor mirror (≥35), which limits the maximal Q-factor of microcavity with semiconductor mirrors. On the contrary, a significant increase in the Q-factor (by about 1.6 times for 5 μm pillar) is possible for micropillar cavity with the proposed hybrid top mirror.

The results of an experimental study of the lasers based on the hybrid micropillar cavity were presented. The minimum laser threshold coincided with a temperature of 130 K, which is close to zero gain to cavity detuning. High thermal stability is achieved due to the non-absorbing mirrors. The lasing action under elevated temperatures (up to 220 K) is due to better vertical mode confinement of the hybrid cavity.

### Acknowledgment

The low temperature studies were partially carried out on the equipment of the large-scale research facilities "Complex optoelectronic stand".

### References

[1] H. Deng, G. L. Lippi, J. Mørk, J. Wiersig, and S. Reitzenstein, "Physics and applications of high-β micro- and nanolasers," *Adv. Opt. Mater.*, vol. 9, no. 19, Jun. 2021, Art. no. 2100415, doi: 10.1002/adom.202100415.

[2] L. Andreoli et al., "Optical pumping of quantum dot micropillar lasers," *Opt. Express*, vol. 29, no. 6, pp. 9084–9097, Mar. 2021, doi: 10.1364/oe.417063.

[3] T. Heindel, J.-H. Kim, N. Gregersen, A. Rastelli, and S. Reitzenstein, "Quantum dots for photonic quantum information technology," *Adv. Opt. Photonics*, vol. 15, no. 3, pp. 613–738, Aug. 2023, doi: 10.1364/aop.490091.

[4] T. Heuser, J. Grose, S. Holzinger, M. M. Sommer, and S. Reitzenstein, "Development of highly homogenous quantum dot micropillar arrays for optical reservoir computing," *IEEE J. Sel. Top. Quantum Electron.*, vol. 26, no. 1, pp. 1–9, Jan. 2020, doi: 10.1109/jstqe.2019.2925968.

[5] P. Senellart, G. Solomon, and A. White, "High-performance semiconductor quantum-dot single-photon sources," *Nat. Nanotechnol.*, vol. 12, no. 11, pp. 1026–1039, Nov. 2017, doi: 10.1038/nnano.2017.218.

[6] S. Kreinberg et al., "Quantum-optical spectroscopy of a two-level system using an electrically driven micropillar laser as a resonant excitation source," *Light: Sci. Appl.*, vol. 7, no. 1, Jul. 2018, Art. no. 41, doi: 10.1038/s41377-018-0045-6.

[7] M. Lermer et al., "High beta lasing in micropillar cavities with adiabatic layer design," *Appl. Phys. Lett.*, vol. 102, no. 5, Feb. 2013, Art. no. 052114, doi: 10.1063/1.4791563.

[8] S. Kreinberg et al., "Emission from quantum-dot high-β microcavities: transition from spontaneous emission to lasing and the effects of superradiant emitter coupling," *Light: Sci. Appl.*, vol. 6, no. 8, pp. e17030–e17030, Feb. 2017, doi: 10.1038/lsa.2017.30.

[9] B. Bahari, A. Ndao, F. Vallini, A. El Amili, Y. Fainman, and B. Kanté, "Nonreciprocal lasing in topological cavities of arbitrary geometries," *Science*, vol. 358, no. 6363, pp. 636–640, Nov. 2017, doi: 10.1126/science.aao4551.

[10] N. Heermeier et al., "Spin-lasing in bimodal quantum dot micropillar cavities," *Laser Photonics Rev.*, vol. 16, no. 4, Feb. 2022, Art. no. 2100585, doi: 10.1002/lpor.202100585.

[11] M. Pflüger, D. Brunner, T. Heuser, J. A. Lott, S. Reitzenstein, and I. Fischer, "Experimental reservoir computing with diffractively coupled VCSELs," *Opt. Lett.*, vol. 49, no. 9, pp. 2285–2288, Apr. 2024, doi: 10.1364/ol.518946.

[12] C.-W. Shih et al., "Low-threshold lasing of optically pumped micropillar lasers with $Al_{0.2}Ga_{0.8}As/Al_{0.9}Ga_{0.1}As$ distributed Bragg reflectors," *Appl. Phys. Lett.*, vol. 122, no. 15, Apr. 2023, Art. no. 151111, doi: 10.1063/5.0143236.

[13] S. A. Blokhin et al., "Design optimization for bright electrically-driven quantum dot single-photon sources emitting in telecom O-band," *Opt. Express*, vol. 29, no. 5, pp. 6582–6598, Feb. 2021, doi: 10.1364/oe.415979.

[14] A. Babichev et al., "Impact of device topology on the performance of high-speed 1550 nm wafer-fused VCSELs," *Photonics*, vol. 10, no. 6, Jun. 2023, Art. no. 660, doi: 10.3390/photonics10060660.

[15] K. Gaur et al., "High-β lasing in photonic-defect semiconductor-dielectric hybrid microresonators with embedded InGaAs quantum dots," *Appl. Phys. Lett.*, vol. 124, no. 4, Jan. 2024, Art. no. 041104, doi: 10.1063/5.0177393.

[16] A. Koulas-Simos et al., "Quantum fluctuations and lineshape anomaly in a high-β silver-coated InP-based metallic nanolaser," *Laser Photonics Rev.*, vol. 16, no. 9, Jul. 2022, Art. no. 2200086, doi: 10.1002/lpor.202200086.

[17] A. Koulas-Simos et al., "High-β lasing in self-assembled photonic-defect microcavities with a transition metal dichalcogenide monolayer as active material," *Laser Photonics Rev.*, just accepted, Jun. 2024, Art. no. 2400271, doi: 10.1002/lpor.202400271.

[18] S. Viciani, M. Gabrysch, F. Marin, F. M. di Sopra, M. Moser, and K. H. Gulden, "Lineshape of a vertical cavity surface emitting laser," *Opt. Commun.*, vol. 206, no. 1–3, pp. 89–97, May 2002, doi: 10.1016/s0030-4018(02)01381-0.

[19] S. Kreinberg et al., "Thresholdless transition to coherent emission at telecom wavelengths from coaxial nanolasers with excitation power dependent β-factors," *Laser Photonics Rev.*, vol. 14, no. 12, Nov. 2020, Art. no. 2000065, doi: 10.1002/lpor.202000065.

[20] E. Schlottmann et al., "Injection Locking of quantum-dot microlasers operating in the few-photon regime," *Phys. Rev. Appl.*, vol. 6, no. 4, Oct. 2016, Art. no. 044023, doi: 10.1103/physrevapplied.6.044023.

**Andrey Babichev** received the Ph.D. degree in condensed matter physics from the Ioffe Institute, St. Petersburg, Russia, in 2014. He has authored or co-authored 182 papers published in refereed journals and conference proceedings. His research interests include semiconductor heterostructures and lasers based on them. Repeated DAAD scholarship as well as Metchnikov scholarship holder. He received the Academia Europaea awards (the Russian Prizes) in 2016. He also admitted the Ioffe Prize in 2020 and 2023.

**Ivan Makhov** was born in Saint Petersburg, Russia, in 1992. He received the master's and Candidate of Science (Ph.D.) degrees from Peter the Great St. Petersburg Polytechnic University, Saint Petersburg, in 2015 and 2021, respectively. He is currently a Researcher at the International Laboratory of Quantum Optoelectronics, National Research University Higher School of Economics, Saint Petersburg. His research interests include quantum dot micro- and macrolasers, optical and electrooptical phenomena in semiconductor-based nanostructures in infrared and terahertz spectral ranges.

**Natalia Kryzhanovskaya** received the Engineering degree from Saint Petersburg State Electrical Engineering University, Saint Petersburg, Russia, in 2002, the Candidate of Science (Ph.D.) degree from the Ioffe Institute in 2005, and the D.Sc. degree from Saint Petersburg Academic University, Saint Petersburg, in 2018. She is currently the Head of the Laboratory of Quantum Optoelectronics, National Research University Higher School of Economics, Saint Petersburg. Her research interests include the study of passive and active optoelectronic devices based on nanostructures.

**Alexey Blokhin** interests include semiconductor heterostructures and its processing technology.

**Yury Zadiranov** interests include semiconductor heterostructures and its processing technology.

**Yulia Salii** interests include semiconductor heterostructures and its processing technology.

**Marina Kulagina** was born in Zlatoust, Russia, in 1960. She received the Graduate degree from the Chemical-Techological Department, Leningrad Forest Academy, Saint Petersburg, Russia, in 1982. She started working at A.F.Ioffe Physico-Technical Institute, Russian Academy of Science, St. Petersburg. She is currently a Research Fellow at the Centre of Nanoheterostructure Physiscs, Ioffe Physical-Technical Institute, St. Petersburg. She has coauthored more than 50 papers. Her research interests include research and development of postgrowth technologies, including photolithography, thin-film deposition, and etching of different semiconductor devices. Mrs. Kulagina received the Russian Academy of Science Diploma for the successful work.

**Mikhail Bobrov** received the Diploma (M.S. equivalent) degree in physics from Peter the Great St. Petersburg State Polytechnic University, Saint Petersburg, in 2014. He has co-authored 74 papers published in refereed journals and conference proceedings. His research interests include the theoretical modeling and characterization of the different optoelectronic devices, including vertical-cavity surface emitting lasers, photodiodes, and single photon sources.

**Alexey Vasil'ev** interests include semiconductor heterostructures and its growth technology.

**Sergey Blokhin** received the Ph.D. degree in semiconductor physics from the Ioffe Institute, Saint Petersburg, in 2006. He has authored or co-authored over 180 papers published in refereed journals and conference proceedings. His current research interests include technology and characterization of III–V semiconductor nanostructures and development of the different optoelectronic devices, including vertical-cavity surface emitting lasers, photodiodes and single photon sources.

**Nikolay Maleev** received the Ph.D. degree in Engineering from the St. Petersburg State Electrotechnical University "LETI", Saint Petersburg, in 1997. He has authored or co-authored over 160 papers and 7 patents. His current research interests include design, modeling and technology of III–V heterostructure devices, including vertical-cavity surface emitting lasers and quantum dot single photon sources.

**Maria Tchernycheva** received the PhD in physics from the University Paris Sud, Orsay (France) in 2005, then worked as a postdoctoral researcher at the Laboratory for Photonics and Nanostructures, CNRS. In 2006, she joined CNRS and is now working as a director of research at the Center of Nanosciences and Nanotechnologies, University Paris Saclay. Her research encompasses the fabrication and testing of novel optoelectronic devices. She published more than 200 articles in international journals, received the Madeleine Lecoq award from the French Academy of Sciences in 2006 and was laureate of the ERC grant in 2014.

**Leonid Karachinsky** received the Ph.D. degree in semiconductor physics from the Ioffe Institute, Saint Petersburg, Russia, in 2004, and Dr. Sci. degree in Engineering from the ITMO University, St. Petersburg, Russia, 2021. Since 2021 he became a Professor and Deputy Director at the Institute of Advanced Data Transfer Systems, ITMO University. He has co-authored over 100 papers. His research interests include semiconductor heterostructures and lasers based on them.

**Innokenty Novikov** received the Ph.D. degree from the Ioffe Institute, Russian Academy of Sciences, Saint Petersburg, in 2005. He has co-authored over 50 papers. His current research interests include development and investigation of the different types of semiconductor lasers and theoretical modeling of optical properties of semiconductor lasers.

**Anton Egorov** received the Diploma degree from the Electrical Engineering Institute in Leningrad in 1987 and the Ph.D. and Dr. Sci. degrees in 1996 and 2011, respectively. He has authored over 410 papers published in refereed journals and conference proceedings. His area of experience includes molecular beam epitaxy; III–V semiconductor heterostructures; GaAs- and InP-based (In,Ga,Al) (As,N) quantum well and quantum dots heterostructures; edge- and vertical-cavity surface emitting lasers (infrared); and III–V heterostructures for microelectronics.